\documentclass[11pt,notitlepage]{article}
\usepackage{euler,times,graphicx}
\usepackage{natbib}
\def\bq{\begin{quote}\begin{singlespace}}
\def\eq{\end{singlespace}\end{quote}}

\newcommand{\footremember}[2]{%
    \footnote{#2}
    \newcounter{#1}
    \setcounter{#1}{\value{footnote}}%
}

\begin{document}
\title{Dark Energy or Modified Gravity?}
\author{Chris Smeenk\footremember{Western}{Department of Philosophy and Rotman Institute of Philosophy, University of Western Ontario, London, ON, Canada. Email: csmeenk2@uwo.ca} and James Owen Weatherall\footremember{UCI}{Department of Logic and Philosophy of Science, University of California, Irvine. Email:  james.owen.weatherall@uci.edu}}

\date{April 20, 2023}

\maketitle
\begin{abstract}
We consider some of the epistemic benefits of exploring ``theory space'' in the context of modifications of general relativity with intended applications in cosmology.  We show how studying modifications of general relativity can help in assessing the robustness of empirical inferences, particularly in inaccessible regimes.  We also discuss challenges to sharply distinguishing apparently distinct directions in theory space.
\end{abstract}

\section{Introduction}

Philosophers often take interpreting theories to be one of their professional obligations, to be discharged by providing an account of what the world would be like if a given theory were true.  But exclusive focus on the theory-world relation, important though that may be, overlooks the insights to be gained by assessing how a theory fits into the space of possible alternative theories.  Physicists routinely consider modifications, reformulations, and generalizations of a given theory in order to gain insight into its structure and viability.  In this paper we will consider some of the reasons for exploring ``theory space'' and the potential benefits to doing so, as exemplified by recent work in cosmology.  

We will argue that exploring theory space is not only common, but also potentially fruitful.  But the case we consider also illustrates some limitations of the approach.  As we discuss, it is often challenging to sharply delineate apparently distinct directions or locations in theory space, on empirical or even conceptual grounds.  We will argue that this undermines some, but not all, of the interpretative benefits of shifting the focus of analysis up one level from theories to theory spaces.  The upshot will be a nuanced view of what a theory space approach can offer.

Our examples of theory space exploration concern modifications to general relativity.  Alternatives to general relativity (GR) have been explored extensively over the last century, with at least three distinct motivations.  First, a modified theory may provide a better path to a successor theory. The application of quantization techniques that have worked for other classical field theories to GR do not yield a perturbatively renormalizable theory. Among the many avenues of research that have been pursued in quantum gravity, one involves modifying classical GR to ease the application of quantization techniques, or to yield a better outcome. Second, the exploration of alternatives has also been used to assess the ``rigidity'' of GR, by showing (for example) that some modifications lead to pathological theories.  It is appealing to seek physical theories that are over-constrained, in the sense that different structural features fit tightly together with little scope for alterations. It is then more plausible that a few physical facts about the nature of gravity, along with general principles, are sufficient to pick out GR as the best (classical) theory of gravity.

Below we will focus primarily on a third motivation: the use of alternative theories to assess the robustness of empirical inferences.  Cosmologists and astrophysics routinely rely on GR to make inferences regarding the types of matter and energy in the universe, and the nature of specific astrophysical systems.  To what extent do the same conclusions follow if GR is replaced with an alternative theory?

The rationale for considering modifications of GR varies for different regimes of applicability. Physicists apply GR to an extremely wide range of physical environments, spanning several orders of magnitude in parameters measuring spacetime curvature and the strength of the gravitational potential.  GR is assumed to fail in the ultraviolet (high-energy, short length-scale) regime, where quantum effects are expected to be relevant.  In cosmology, the target systems are too big for GR to fail for this reason. There are no comparably persuasive theoretical arguments that suggest a ``maximum length scale'' for GR. Yet cosmology involves an enormous extrapolation from the length scales where GR has been most rigorously tested -- the scale of the solar system, or even smaller systems such as binary black holes.  

Extrapolating over some $14$ orders of magnitude is surely enough to make cautious empiricists wary.  Active exploration of alternatives to GR at large length-scales started in earnest in the early 2000's in response to the discovery that the universe's rate of expansion is accelerating.  In light of these observations, GR (along with standard cosmological assumptions) yields the striking conclusion that the vast majority of mass-energy in the universe comes in the form of an effective cosmological constant ($\Lambda$) or a type of matter that mimics its effects, ``dark energy.''  Though classical general relativity could accommodate an accelerating universe without modification, cosmologists have generally been relucant to accept such a large $\Lambda$-like contribution.  Instead, many have considered it natural to see whether a cosmological constant can be avoided by modifying GR at cosmological scales.

Here we focus on the benefits and challenges to exploring theory space as illustrated by this case of ``dark energy'' research, broadly construed.  We begin with a brief overview (in \S \ref{Sec:Lambda}) of contemporary dark-energy phenomenology, with an emphasis on the role of assumptions of different types (including GR, but also cosmological assumptions) in supporting the inference to a large dark energy contribution. In \S \ref{Sec:Closing the Loop?} we discuss three different responses to these phenomenological results:  treating $\Lambda$ as a true constant, treating ``dark energy'' dynamically as a new form of matter, and modifying GR. We suggest that cosmologists' exploration of these possibilities can be understood as a version of ``closing the loop'', as described by George \citet{smith2014closing}.  This epistemic strategy, we suggest, has the potential to provide a powerful evidence when it succeeds.  But in this case, we argue, it has not succeeded. Instead, cosmologists are to some degree victims of their own success, with no opportunity to sharply distinguish between the theoretical options under consideration through further comparisons with observations.  

We then turn in \S \ref{Sec:Foundations} to question whether there is a clear contrast to be drawn at the foundational level between modified gravity and dark energy scenarios, and we critically assess one proposal for drawing that distinction.  Finally, in the conclusion we return to assess the implications of these difficulties for theory space analysis of the sort considered here.  It is feasible to assess the sensitivity of various empirical inferences to theoretical assumptions of different modifications without also classifying the modifications.  The assessment of the plausibility of the modifications, however, or taking the phenomenology as a guide towards a successor theory, does require resolving the question in our title.  

\section{Measuring $\Lambda$}
\label{Sec:Lambda}

The introduction of a cosmological constant $\Lambda$ was, in effect, the first modification of GR:  Einstein added the term two years after discovering his now-eponymous equation. Even for those wary of Einstein's motivations, there are no compelling physical grounds to set its value to zero, as opposed to leaving it as a parameter to be determined by observations.  Many mid-century treatments of relativistic cosmology set aside $\Lambda$, perhaps agreeing with Einstein's regretful assessment that introducing it had been a mistake.  Yet there have always been hints that $\Lambda$ should not be ignored.  Early cosmological models faced an age crisis:  some astrophysical objects (globular clusters, stars) appeared to be significantly older than the universe itself.  A non-zero $\Lambda$ severs the connection between the current expansion rate and age of the universe responsible for this conflict.  Several cosmologists preferred a model with flat spatial sections --- such that the total matter-energy density sums to the so-called critical density.  As observational estimates of matter density increased in accuracy, it became increasingly clear that reaching critical density would require a large contribution from $\Lambda$.   These were two of the more prominent reasons cosmologists gave for taking $\Lambda$ seriously, and there are several others of varying significance.

This situation changed dramatically in the late 90s, as cosmologists developed a compelling observational case that $\Lambda$ accounts for roughly $70 \%$ of the total mass-energy density of the universe --- in other words, $\Omega_{\Lambda} \approx .7$.\footnote{We do not have space to cite original papers here; see, e.g., \citet{peebles2020cosmology} for an overview. The density of a flat model is given by  $\rho_c = \frac{3}{8\pi G} \left(H^2 - \frac{\Lambda}{3}\right)$, where $H$ is the Hubble constant. The density parameters for different types of matter are then ratios with respect to the critical density, $\Omega_i = \frac{\rho_i}{\rho_c}$, with $\Lambda$ set to zero; for dark energy, $\Omega_{\Lambda} = \frac{\Lambda}{3H^2}$.} %; and $\Omega_k = - \frac{k}{R^2 H^2}$ (where $k= \{-1,0,+1\}$ for negative, flat, and positive curvature.} 
During this period, cosmology transitioned from providing qualitative explanations of observed large-scale features of the universe to constraining parameters appearing in the ``standard model'' with increasing precision.  %The parameters characterize, for example, the expansion rate (the Hubble constant $H$), and the density of different types of matter (such as $\Omega_m$, the total matter contribution, including baryonic and dark matter, and $\Omega_{\Lambda}$). 
Accepting the standard model allows cosmologists to bring a wide array of observations to bear to measure parameters, including the Hubble constant, $H$, and the density of different types of matter ($\Omega_m$, the total matter contribution, including baryonic and dark matter, and $\Omega_{\Lambda}$).   Cosmologists have clarified what types of observations provide the most precise constraints on different parameters and have also pursued sensitivity analysis, identifying the assumptions required to link observations to a parameter (or set of parameters) and then assessing to what extent this link is robust to variations.    

Three types of observations provide the strongest constraints on the value of $\Omega_{\Lambda}$. The first, going back to Hubble's seminal observations of an approximately linear relationship between redshift and magnitude, takes galaxies as tracers of cosmic expansion.  Granting that the universe at large scales can be approximately described by the simple homogeneous and isotropic models (the FLRW models), the dynamical equations support inferences from the redshift-magnitude relationship of ``standard candles'' to the densities and types of matter present.  Ordinary matter leads to decelerating expansion --- that is, $\ddot{R}(t) < 0$, where $R(t)$ is the scale factor.  The probative value of this kind of observation increases with distance, and starting in the 90s astrophysicists exploited supernovae to extend to larger distances.  Although they expected this to yield more precise measurements of deceleration, they instead discovered that the expansion rate appears to be accelerating.  This implies that the dominant dynamical contribution (at late times) must come from a $\Lambda$-like source, needed to yield $\ddot{R}(t) > 0$.  The two teams pursuing this line of observations both concluded that the ``best fit'' FLRW model has $\Omega_m \approx .3$ and $\Omega_{\Lambda} \approx .7$. 
\begin{figure}
    \centering
    \includegraphics[height=5.5cm]{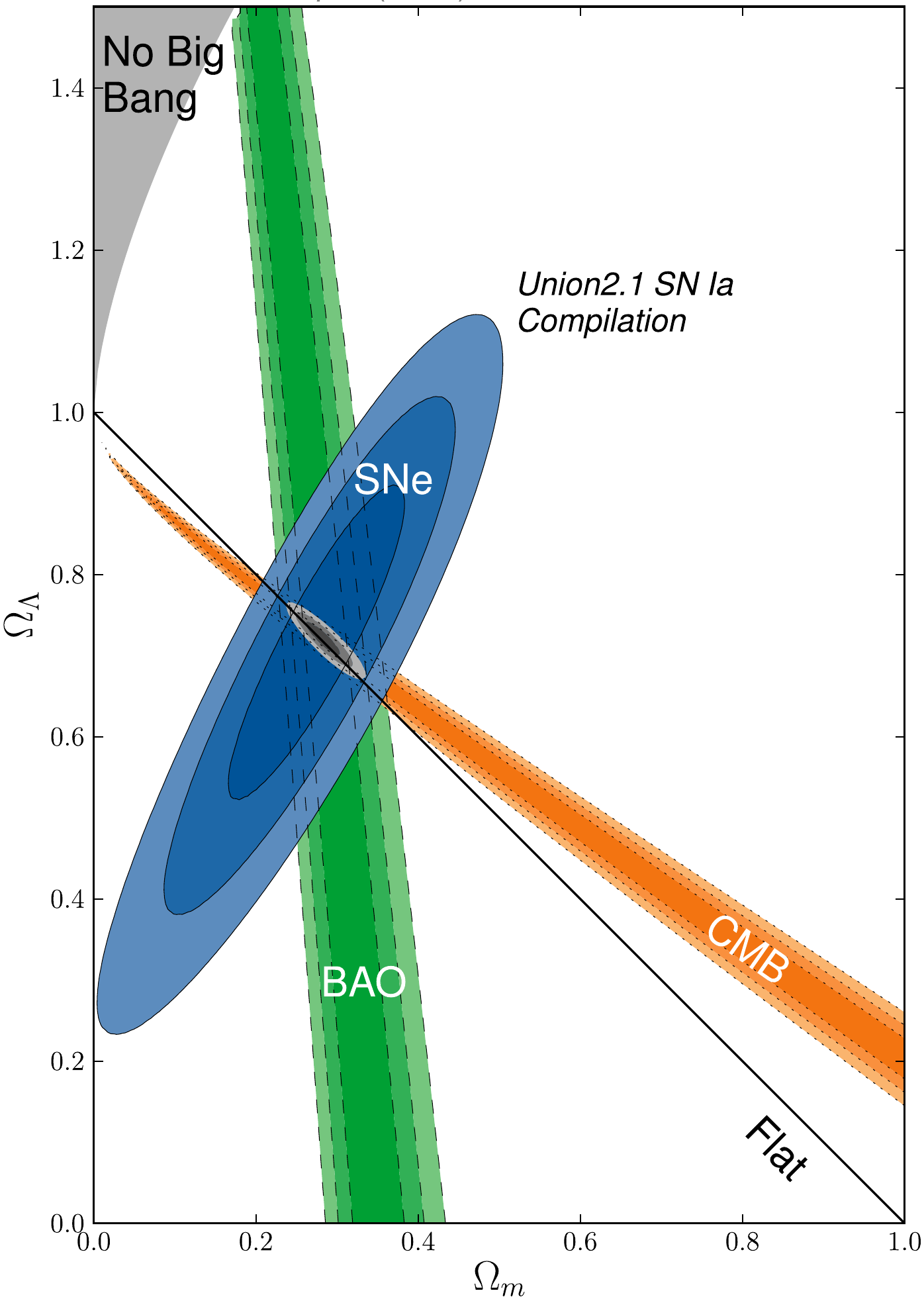}    \caption{The contours show parameter estimates at standard confidence levels (only statistical error) for measurements from baryon acoustic oscillations (BAOs), supernovae (SNe), and the CMB \citep[Figure 5 from][]{suzuki2012hubble}}
    \label{fig:ConstraintsLambda}
\end{figure}

This discovery came as a surprise to both teams, but two other lines of reasoning had already suggested a large value of $\Lambda$. Simulations of structure formation developed in the 90s seemed to require a flat universe with close to critical density.  Given observational upper bounds on the total matter density, it was fairly routine to simply assume a large value of $\Lambda$.  This was at best a qualitative constraint.  Subsequent work has leveraged an understanding of structure formation to generate constraints on the underlying cosmological model, through the study of baryon acoustic oscillations.  
Later work in structure formation has generated precise constraints, primarily through the sutdy of baryon acoustic oscillations.  
Cosmological perturbations excite sound waves in the primordial plasma in the early universe, whose propagation comes to an abrupt end at recombination.  The maximum distance the oscillations can travel during this time sets a characteristic length scale.  This length scale can be measured directly from the acoustic peaks in the CMB (cosmic microwave background).  But granting a theoretical account of subsequent phases of structure formation, the same characteristic length scale can be measured through observing the large-scale clustering of galaxies at later times.  These oscillations provide cosmologists with a ``standard ruler,'' and observations of this length scale at different cosmological epochs constrains the expansion rate of the universe and the cosmological parameters.  

Observations of the CMB also suggested the need for a large $\Lambda$. The angular position of the first acoustic peak in the power spectrum of temperature fluctuations in the CMB, for example, reveals that the universe at recombination has a flat spatial geometry. This implies that the total matter-energy density is close to the critical density ($\Omega_{total}=1)$, but only weakly constrains individual contributions from $\Omega_m$ and $\Omega_{\Lambda}$.  By the mid-90s, however, several cosmologists made the case that other lines of evidence lead to an upper bound on $\Omega_m$, forcing the introduction of a large contribution from $\Omega_{\Lambda}$ \citep[e.g.,][]{OstrikerSteinhardt1995}.

Contour plots such as figure \ref{fig:ConstraintsLambda} display confidence intervals for parameter estimates from these different types of observations.  Each of these three methods has limitations --- the CMB and SNe measurements have significant degeneracy in the values of $\Omega_m$ and $\Omega_{\Lambda}$, whereas BAOs place tight constraints on $\Omega_m$ but have little to say about $\Omega_{\Lambda}$. Yet these degeneracies extend in different directions in parameter space, such that the three methods in conjunction do yield tight constraints on $\{\Omega_m,\Omega_{\Lambda}\}$. Furthermore, each of these methods employs distinctive observational techniques and rests on different physical and cosmological assumptions.  All three measurements rely on aspects of general relativity (to draw inferences from observed motions to the underlying sources) along with the approximate validity of FLRW models.  They differ in several other respects, ranging from the kinds of objects studied to the physical and astrophysical details relevant to turning observations into parameter constraints.  As \citet{Harper2011} noted, cosmologists have taken this consilience of inductions between three different methods to boost confidence in the observational case for dark energy.      

 Cosmologists have discovered something remarkable about the universe. Yet what is it?  The phenomenology summarized above is compatible with three answers.    First, treat $\Lambda$ as a free parameter whose value is to be determined by observations, and regarded as a contingent brute fact.  On this view, the main question is then whether the different lines of evidence determine a consistent value of $\Lambda$.  Second, attribute this phenomenology to a new type of mass-energy, ``dark energy,'' characterized by the fact that it mimics $\Lambda$ in certain regimes.  Hypothesized dynamics for dark energy could then lead to a richer phenomenology, differing from that of a true cosmological constant.  Third, note that the inferences above all rely on GR applied at large length scales, in conjunction with cosmological assumptions.  Modifying GR in this regime may make it possible to account for the phenomenology without dark energy or a true cosmological constant.

\section{Closing the Cosmological Loop?}
\label{Sec:Closing the Loop?}

Why, if well-established theory apparently accounts for all known observations, would anyone set aside the cosmological constant in favor of more exotic alternatives?   Physicists are clearly not satisfied with merely measuring $\Lambda$, and have introduced different possible ways of filling out the underlying physics.  

One compelling answer that explains the appeal of doing so draws on George Smith's account of the methodology of celestial mechanics \citep{smith2014closing}.  Over two centuries, astronomers iteratively developed more detailed accounts of the solar system by identifying robust physical sources for observational discrepancies, and then adding these new sources back into an enriched theoretical model.  There are three features of Smith's analysis that are particularly salient.  First, Smith emphasizes the need to have sufficient control over idealizations appearing in the description at any given stage of inquiry: the idealizations should identify regularities that would be exact in precisely specifiable (albeit counterfactual) circumstances.  Second, measurements of fundamental quantities should remain stable even as the model undergoes further refinement.  Third, the addition of robust sources have a variety of further downstream impacts in the subsequent modeling, that extend well beyond the resolution of the initial discrepancy. Smith argues that repeatedly closing the loop, by finding robust sources and adding them to the model, and pursuing the further implications of the enriched model, provides compelling evidence in favor of both the theory and the individual details incorporated at each successive stage.   

It is clear why treating $\Lambda$ as a true constant is unappealing:  it is sterile, and fails to generate further consequences that can be pursued through theory or observational programs.  It represents a dead end rather than a step towards further iterative refinements.  It is much more appealing to assume that there is a robust physical source hiding under the facade of a true constant. \citet{peebles2003cosmological} for example, note in the abstract of their review article that: ``Physics welcomes the idea that space contains energy whose gravitational effect \emph{approximates} that of Einstein's cosmological constant, $\Lambda$; today the concept is termed dark energy or quintessence'' (our emphasis, p. 559).  It is crucial that the effects of dark energy only \emph{approximate} that of a cosmological constant, so that observational programs can isolate the differences and begin the process of iteratively learning more about the physics of dark energy.  

\begin{figure}
    \centering
    \includegraphics[height=5.0cm]{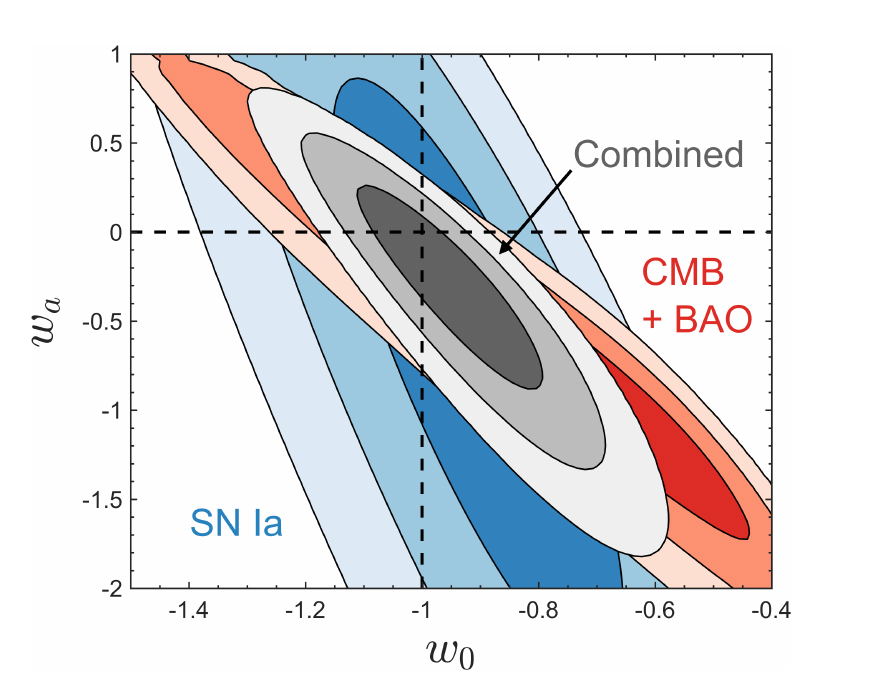}    \caption{The contours show parameter estimates at standard confidence levels for measurements of $(w_0,w_a)$ from CMB, BAO, and SN Ia \citep[Figure 9 from][]{huterer2017dark}}
    \label{fig:Constraintsw}
\end{figure}

 One class of dark energy models illustrates how the physics of dark energy could be revealed through cosmological observations.  Quintessence  models treat dark energy as a self-interacting scalar field, weakly coupled to other fields.  There are two significant constrasts between these models and a true cosmological constant: the evolution of the scalar field with cosmic time generates a time-variation in the effective $\Lambda$, whereas a true $\Lambda$ is fixed once and for all, and the field's effective equation of state can vary (both as a function of space and / or time).  Cosmologists usually characterize these contrasts in terms of two parameters.  The equation of state for a perfect fluid is given by $w = \frac{p}{\rho}$, where $p$ is the pressure and $\rho$ the energy density.  The first parameter $w_0$ reflects the current measured value of the equation of state; a second parameter $w_a$ characterizes its dependence on time (red-shift).  For a true cosmological constant, $w_0 = -1$ and $w_a = 0$; scalar field models are expected to lead to departures on large length and/or time scales.  
  
 After two decades of dedicated work, cosmologists have yet to find compelling evidence for such variations; the boring answer --- a true cosmological constant, indicated by the dashed lines in figure \ref{fig:Constraintsw} --- still fits the available data.  As far as we are aware, there are at present no cosmological observations incompatible with taking $\Lambda$ to be a true constant.  To put it in slightly different terms, requiring that dark energy reproduces the phenomenology described above renders any new features inaccessible.  The process of iterative development of dark energy models cannot even get started.   

There are also foundational reasons for rejecting a true cosmological constant.  Indeed, many cosmologists fail to consider the possibility that $\Lambda$ could be regarded as a universal constant of nature, and instead immediately identify it with vacuum energy density.  In quantum field theory, the stress-energy tensor for vacuum energy density has the same form as a cosmological constant term.  Conventional wisdom goes beyond this formal analogy to take the cosmological constant as nothing but vacuum energy; doing so leads to the ``cosmological constant problem'' given the remarkable difference between the cosmological constraints on $\Lambda$ discussed above and calculations of vacuum energy density in QFT.\footnote{We cannot pursue the topic further here, but see \citet{koberinski2022lambda,wallace2022quantum} for recent philosophical discussions.}  For several decades, physicists have sought new symmetries or other mechanisms that would provide a physical argument that vacuum energy density vanishes, which would imply (granting the presumed identity) a vanishing cosmological constant.  The search continues, but even without a compelling argument along these lines cosmologists have treated the first option as a non-starter.

Finally we turn to the possibility that modifications of GR account for the phenomenology described above. The measurements of $\Lambda$ all depend on the dynamics of GR and the approximate validity of the FLRW models.  Several cosmologists pursued the plausible suggestion that changing the long-range behavior of gravity could suffice to explain accelerating expansion (and other observations). One of the earliest proposals was based on $f(R)$ theories, which replace the Einstein-Hilbert Lagrangian density, $\mathcal{L}_{EH} = kR$, where $k$ is a constant and $R$ is the scalar curvature, with some function of the scalar curvature, $\mathcal{L} = f(R)$. \citet{hu2007models}, for example, constructed an $f(R)$-theory that accounts for accelerated expansion without a cosmological constant or dark energy.  The model can only reconcile this cosmological behavior with the constraints imposed by matching the success of GR on much solar-system scales by introducing a mechanism to segregate the different scales.  It is remarkable that the success of GR on solar-system scales places such tight constraints on GR at cosmological scales.  But constructions like \citet{hu2007models}, and all other models we are aware of that match solar system constraints, dramatically limit the possibility of further tests of the proposed novel physics.  Modified gravity scenarios are victims of GR's success:  the downstream consequences of the proposed new physics is simply not within empirical reach.

\section{Foundations}
\label{Sec:Foundations}

Thus far, we have argued that there are in-principle challenges to empirically distinguishing modified gravity from dark energy scenarios or a true cosmological constant.  In doing so, we have taken for granted that there is a clear conceptual difference.  But as we will presently argue, it is not clear that a firm conceptual distinction between modified gravity and dark energy is possible, either.  Space constraints forbid a complete discussion of this issue, but some brief remarks will point towards the difficulties.

To begin, we acknowledge that some theories appear to be unambiguous examples of modified gravity.  Consider, for instance, $f(R)$ theories \citep{fRReview}.  Such theories generally lead to higher-order generalizations of Einstein's equation, but they do not lead to new degrees of freedom that might be taken for matter fields.\footnote{That said, even $f(R)$ theories can lead to ambiguities, since they can generally be re-expressed in a form that does introduce new degrees of freedom, by performing a conformal transformation on the metric.  We set this issue aside for now.}

But other theories are more ambiguous.  Take Horndeski theories.  These are theories described by the Horndeski Lagrangian density, which is the most general Lagrangian density depending only a Lorentzian metric and scalar field for which the Euler-Lagrange equations are second order \citep{horndeski}.  Horndeski theories have received considerable attention from theoretical cosmologists over the past decade, as they encompass a very broad class of ``reasonable'' candidates for reproducing dark energy phenomenology via a new dynamical field \citep{modGravReview}.  Studying the entire class of theories together, as special cases of a single very general Lagrangian density with several unknown functions, permits one to place constraints on many possible theories at once and rule out large swaths of theory space efficiently.  This is a potentially powerful form of theory-space analysis.

But the generality of the Horndeski Lagrangian also introduces conceptual puzzles.  Among the Horndeski theories are clear cases of dark energy candidates, including quintessence \citep{Caldwell+etal}.  But also among the Horndeski theories are classic cases of modified gravity theories, such as Brans-Dicke theory \citep{Brans+Dicke}, which modifies general relativity by replacing the gravitational constant $G$ by a dynamical field $\varphi$ governed by its own dynamics.  And one can also find hybrid theories that include elements of both. Indeed, even Brans-Dicke theory, viewed from this perspective, might be seen as an intermediate case, since the dynamical gravitational constant in that theory also contributes to the stress-energy tensor and, in particular, can exchange energy and momentum with matter.  This is so despite the fact that Brans and Dicke themselves intended it as a ``modified relativistic theory of gravitation'' (abstract).

Cosmologists' rare attempts to sharpen this distinction do not hold up under scrutiny.\footnote{\citet{martens2020dark} wrestle with the modified gravity / matter distinction in the context of dark matter, and propose a detailed taxonomy of different approaches.  Space constraints prohibt detailed engagement with their approach, though we suggest that the issues are even more thorny in the dark energy case.}  Consider, for instance, the relatively infuential proposal due to \citet{JoyceDEModifiedGrav}, who argue that as a pragmatic matter, we should understand modified gravity theories to be ones that lead to a ``fifth force''.  To make this precise, they invoke a version of the strong equivalence principle (SEP).  The idea is that dark energy theories will be ones that satisfy the SEP, whereas modified gravity theories will not.

This is a compelling proposal, but the details do not work.  \citet{JoyceDEModifiedGrav} propose their own statement of the SEP, which is the requirement that all massive bodies, including compact bodies such as black holes, follow geodesics (103).  But, as has been observed many times in the literature on geodesic motion in general relativity, there is no canonical or unambiguous way to associate an extended body with a curve in general relativity, and so ``geodesic motion'' is not well defined \citep[cf.][]{Geroch+Weatherall}. At best this idea makes sense in a small-body limit.  But even if one does consider the small-body limit, at first-order in the size of the body, one expects deviations from geodesic motion due to finite body effects, such as angular momentum \citep{Gralla+Wald}.  Thus, even in general relativity one cannot expect compact bodies such as black holes to ``follow geodesics'', except in some approximate or highly idealized sense.  It seems geodesic motion for compact bodies cannot clearly distinguish theories that otherwise approximate general relativity. 

One might hope that other statements of the strong equivalence principle would do better.  It is possible that they would.  But several authors have recently argued that there are multiple such statements in the literature that do not generally agree; and in any case, many of them are not sufficiently clear to apply in cases where there is any actual dispute \citep{Fletcher+Weatherall}.  Moreover, other standard statements, which tend to concern the ``locally special relativistic'' character of matter field equations, do not appear to capture \citet{JoyceDEModifiedGrav}'s intuition that ``fifth force'' effects are characteristic of modified gravity.  

\section{Conclusion}

Physicists often devote serious time and attention to alternatives to well-established theories, even when empirical considerations do not force them to do so.  Examples include the practice of phenomenological model building in particle theory, some work in the foundations of quantum theory, and, as we have discussed here, the study of modified gravitational theories in dark-energy cosmology (among others).  These activities can play many epistemic and other roles, and the practice deserves further scrutiny from philosophers of science.  In this paper we have focused on the promise and perils of one role this sort of activity can play in modern physics, which is to identify potentially fruitful opportunities for theory-mediated measurements that, under the right circumstances, can provide a particularly strong form of evidence.  In this we took inspiration from the ``Closing the Loop'' methodology described by George Smith, who has argued that an iterative process of identifying robust physical sources for unexpected phenomena and then studying the downstream effects of those sources was key to developing the rich evidence for Newtonian gravitational theory that was accumulated between the 1680s and 1900.  

But alas, while we maintain this strategy is both powerful in principle and successfully used in other cases in cosmology, it has not yielded results in the present case, for reasons we suggest are informative about both cosmology and the method. In the case of dark energy research, it is increasingly clear that both modifications of general relativity and exotic matter theories proposed to mimic a cosmological constant are highly constrained by established physics on other scales. Thus they must be carefully tuned so that their only downstream effects concern precisely the phenomena that they were introduced to explain.  For this reason, while some of these proposals would count as potentially robust physical sources for cosmic acceleration, that robustness cannot be established via their effects on other systems or in other regimes. The loop cannot close.  To make matters worse, we argued, even conceptually it turns out that it is difficult to sharply distinguish proposed sources of apparently different types.  This means that even qualitative proposals, the details of which might be filled in in different ways, do not lead to clear observational signatures that might suggest that one path, rather than another, is the most fruitful way forward.  The result is that every proposal that is viable mimics every other proposal that is viable, empirically and conceptually.

Of course, this does not mean that theory space exploration is a useless enterprise, or that it was a bad idea in this particular case.  Instead we take this discussion to show one reason that physicists might study theories with minimal empirical justification, as part of a broader research strategy---but also to highlight the failure modes for that kind of approach, since the main example of a similar methodology in the literature is one of extraordinary success.

\bibliographystyle{psalike}
\bibliography{PSA2022.bib}

\end{document}